\documentclass[10pt,twocolumn]{article}

\usepackage[T1]{fontenc}

\usepackage[margin=0.75in]{geometry}

\usepackage{tabularx}
\usepackage{array}
\usepackage[most]{tcolorbox}
\usepackage{xcolor}
\usepackage{multirow}
\usepackage{amsmath}
\usepackage{amssymb}
\usepackage{makecell}
\usepackage{booktabs}
\usepackage{caption}
\usepackage{subcaption}
\usepackage{graphicx}
\usepackage{url}
\usepackage{hyperref}
\usepackage{tikz}
\usepackage{listings}
\usepackage{changepage}
\usepackage{dblfloatfix}
\usepackage{placeins}
\usepackage{amsthm}
\usepackage[numbers,sort&compress]{natbib}
\usepackage{adjustbox}
\usepackage[compact]{titlesec} 
\titleformat{\section}
  {\normalfont\large\bfseries} 
  {\thesection}{1em}{}

\titleformat{\subsection}
  {\normalfont\normalsize\bfseries} 
  {\thesubsection}{1em}{}

\titleformat{\subsubsection}
  {\normalfont\small\bfseries} 
  {\thesubsubsection}{1em}{}

\newcommand*\circled[1]{\tikz[baseline=(char.base)]{
  \node[shape=circle,fill,inner sep=1pt] (char) {\textcolor{white}{#1}};}}

\newtheorem{_smalldefn}{Definition}
\newenvironment{smalldefn}
  {\begin{_smalldefn}\small}
  {\end{_smalldefn}}

\usepackage[labeled]{multibib}
\newcites{R}{Repositories}

\makeatletter
\providecommand{\@citeR}[2]{[#1\if@tempswa , #2\fi]}
\renewcommand{\@citeR}[2]{R[#1\if@tempswa , #2\fi]}
\renewcommand{\@biblabelR}[1]{R[#1]}
\makeatother

\begin{document}

\title{Recipe for Discovery: A Pipeline for Institutional Open Source Activity}

\author{
\and Juanita Gomez\\
\texttt{jgomez91@ucsc.edu}
\and
Emily Lovell\\
\texttt{elovell@ucsc.edu}
\and
Stephanie Lieggi\\
\texttt{slieggi@ucsc.edu}
\and
\and
Alvaro A. Cardenas\\
\texttt{alacarde@ucsc.edu}
\and
James Davis\\
\texttt{davis@cs.ucsc.edu}
\and
\\
\parbox{\textwidth}{\centering
University of California, Santa Cruz\\
Santa Cruz, California, USA}
}

\date{\today} 
\maketitle

\begin{abstract}
Open source software development, particularly within institutions such as universities and research laboratories, is often decentralized and difficult to track. Although academic teams produce many impactful scientific tools, their projects do not always follow consistent open source practices, such as clear licensing, documentation, or community engagement. As a result, these efforts often go unrecognized due to limited visibility and institutional awareness, and the software itself can be difficult to sustain over time. 

This paper presents an end-to-end framework for systematically discovering and analyzing open source projects across distributed academic systems. Using ten universities as a case study, we build a pipeline that collects data via GitHub's REST API, extracts metadata, and predicts both institutional affiliation and project type (e.g., development tools, educational materials, websites, documentation). Applied across the ten campuses, our method identifies over 200,000 repositories and collects information on their activity and open source practices, enabling a deeper understanding of institutional open source contributions.

Beyond discovery, our framework enables actionable insights into institutional open source practices, revealing patterns such as missing licenses or limited community engagement. These findings can guide targeted support, policy development, and strategies to strengthen open source contributions across academic institutions.

\end{abstract}
\maketitle

\section{Introduction}

Open source software plays a central role in modern software engineering as the foundation of critical infrastructure and scientific discovery. Despite its impact, understanding where and how open source is produced remains a challenge, particularly when it is developed by decentralized institutions such as universities and research laboratories. In these settings, development is often distributed across accounts with inconsistent practices such as missing licenses or limited documentation. As a result, many valuable projects go unrecognized and can be difficult to sustain.

Institutions worldwide are increasingly investing in initiatives to support and strengthen their open source ecosystems. Programs such as the Institutional Innovation Grapher at the University of Texas at Austin \cite{UTOSPOGrapher}, the OSPO Database Development Playbook at the University of Wisconsin \cite{GoringUWProject}, the Stanford Open Source Registry \cite{StanfordOSRegistry}, the George Washington University Open Source Project Registry \cite{gwu_project_registry}, the Janelia OSSI initiative at the Howard Hughes Medical Institute \cite{JaneliaOSSIProjects}, and France’s national research software catalog \cite{MinEduResearchCatalogue} demonstrate a growing commitment to creating tools that help discover, catalog, and showcase institutional open source projects. By mapping these repositories, institutions can provide support to researchers, improve infrastructure and sustainability, and increase the visibility and impact of their open source contributions.

Unlike software developed within companies or centralized teams, academic open source often originates from small university groups or individual researchers. These projects lack a unified platform, naming convention, or metadata standard, and while GitHub is widely used for collaboration, its vast scale makes discovery and analysis difficult. Institutional information in contributor profiles is often inconsistent or missing, repositories may include forks or class assignments rather than active research software, and contributors frequently use multiple profiles or change affiliations. As a result, simple keyword searches (e.g., “UC Davis”) can return thousands of ambiguous results, making manual filtering impractical.

In this paper, we present a scalable approach for systematically discovering and analyzing academic open source repositories. Our method combines GitHub’s REST API with a filtering pipeline that integrates large language models (LLMs) to infer remove false positives and categorize projects by type. Beyond merely cataloging repositories, the framework allows detailed examination of project characteristics, revealing trends such as absent licenses or inconsistent community standards (e.g., contributing guides and templates). These findings highlight opportunities for institutions to provide targeted support, enhancing software sustainability, compliance, and community engagement. 

We use the University of California (UC) system as a case study for evaluating our pipeline because it represents one of the world’s largest public research university systems comprising campuses with varying sizes, disciplinary strengths, and levels of open source activity. Applied across the ten UC campuses, our pipeline identifies 236027 repositories. Given the global impact of the UC system and notable open source projects originating from its campuses (e.g., Jupyter from UC Berkeley), our results offer both local and broader insights for strengthening institutional open source practices. Our contributions include: 


\begin{itemize}
    \item We introduce a scalable method for discovering open source repositories affiliated with institutions (including academic and research organizations) by defining a \textbf{repository signature} based on repository and contributor metadata.
    
    \item We analyze the results of mining the ten UC campuses, producing actionable insights into open source practices, including patterns related to licensing, documentation, and community standards that can guide support and policy.
    \item We release the code for repository scraping, filtering and analysis as an open source tool, enabling replication and further research. The code is available at: \url{https://anonymous.4open.science/r/repofinder-E250}.
\end{itemize}

\section{Related Work}

There is growing interest in discovering and analyzing open source software repositories to better understand their purpose, usage, and practices. For \textbf{discovery}, some universities have developed tools such as the Institutional Innovation Grapher \cite{UTOSPOGrapher}, which identifies GitHub accounts that mention a university in their bios. However, this method has limitations, as many developers do not explicitly mention institutional affiliations. Similarly, the University of Wisconsin Playbook \cite{GoringUWProject} aims to discover open source contributions but does not automate affiliation detection. Schwartz et al. \cite{schwartz2024} present a more comprehensive effort by combining GitHub keyword searches with university website scraping and LLM-based filtering to identify research software engineering (RSE) repositories; however, affiliation is not directly studied.

For \textbf{categorization}, previous work has addressed several angles. One of the most common, is distinguishing software engineering projects from other non-research projects. Munaiah et al. \cite{Munaiah2017} implemented a tool called Reaper to classify engineered software projects using a combination of score-based heuristics and random forest classifiers. Pickerill et al. \cite{Pickerill2020} extended this idea with PHANTOM, a time-series clustering method based on Git logs. 

Many efforts focus on categorizing repositories by topic and functionality \cite{DiRocco2020}, \cite{DiSipio2020}, \cite{Wang2012} \cite{Widyasari2023}, \cite{izadi2021}. Sharma et al.\cite{Sharma2017} proposed automatic cataloging based on topic modeling of README files, while Zhang et al. \cite{zhang2021} used keyword-driven hierarchical classification to assign repositories to topic categories. Aslam et al. \cite{Aslam2022} built a multilingual source code classifier using machine learning, and Linares-Vasquez et al.\cite{LinaresVasquez2014} classified software by application domain using (API) calls from third-party libraries. Embedding-based models have also been explored: Bharadwaj et al. ~\cite{Bharadwaj2022} applied BERT-style models to classify GitHub issues, while Izadi et al.~\cite{Izadi2023} used semantic knowledge graphs to improve topic recommendation.

Regarding \textbf{analysis}, prior work has explored repository practices and the sustainability of open source projects. For example, Ehls et al.~\cite{Ehls2017} analyze sources of project failure to identify risks in open source development, while Alami et al.~\cite{Alami2024} apply sustainability metrics to evaluate how different aspects of open source communities affect software quality. Surveys such as the 2024 Open Source Survey at UW-Madison \cite{UWOS2024} provide insights into usage patterns and opportunities to improve institutional open source practices.

GitHub mining at scale has also been a focus of previous work. Kalliamvakou et al.~\cite{Kalliamvakou2014} highlighted challenges in GitHub data quality and usage patterns. Ma et al.~\cite{Ma2019} introduced World of Code, an infrastructure for mining global VCS data and exploring interdependencies. Cosentino et al.\cite{Cosentino2017} conducted a systematic mapping study of related GitHub research.

Although these works advance methods for repository discovery, classification, and analysis of open source practices, they generally focus on project characteristics, topic modeling, or development activity, without directly addressing institutional affiliation; or they are limited in scope, focusing on individual projects, small samples, or survey-based data, making it challenging to generalize findings at scale or systematically connect them to institutional contexts.

A recent systematic mapping study~\cite{balla2025} surveys 43 works on repository classification and notes that, while many papers cite "discoverability" as a motivation, none explicitly integrates institutional affiliation, project categorization, and actionable analysis at scale. Our work addresses this gap by developing an end-to-end pipeline that systematically identifies institutionally affiliated repositories, classifies them by project type, and analyzes their practices to generate actionable insights. This integrated approach provides institutions with the means to understand and improve their open source ecosystems at scale.

\section{Methodology}

We now describe our methodology for identifying and characterizing open source projects on GitHub. We first explain how we discover candidate repositories using a high recall search strategy across repositories, organizations, and users. We then detail the filtering step used to remove false positives. Finally, we describe how the remaining repositories are further categorized by project type to facilitate the analysis of the academic open source landscape.

\subsection{Discovery}

As the starting point for our data collection, we focus on GitHub because of its dominance in open source development, extensive API support, and widespread adoption. 
To clarify our data collection process, we first summarize the basic structure of a GitHub repository and the types of metadata available through the GitHub API. A \textbf{repository} is the central unit of collaboration on GitHub, containing a codebase, documentation, and configuration files. Each repository includes a number of descriptive \textbf{attributes}, such as its name, description, README file, and associated tags. Repositories may also include \textbf{community files}, like a code of conduct, contributing guidelines, and security policies, files that tend to signal project maturity and openness to external contributors. \textbf{Engagement indicators} such as stars, forks, downloads, and subscriber count reflect how active and visible a project is within the open source ecosystem. Furthermore, each repository is owned by an individual or an \textbf{organization} (a special GitHub account for groups), and includes a list of \textbf{contributors} who have committed code, often with metadata such as their username and affiliation.

\begin{figure}[h]
  \includegraphics[width=0.5\textwidth]{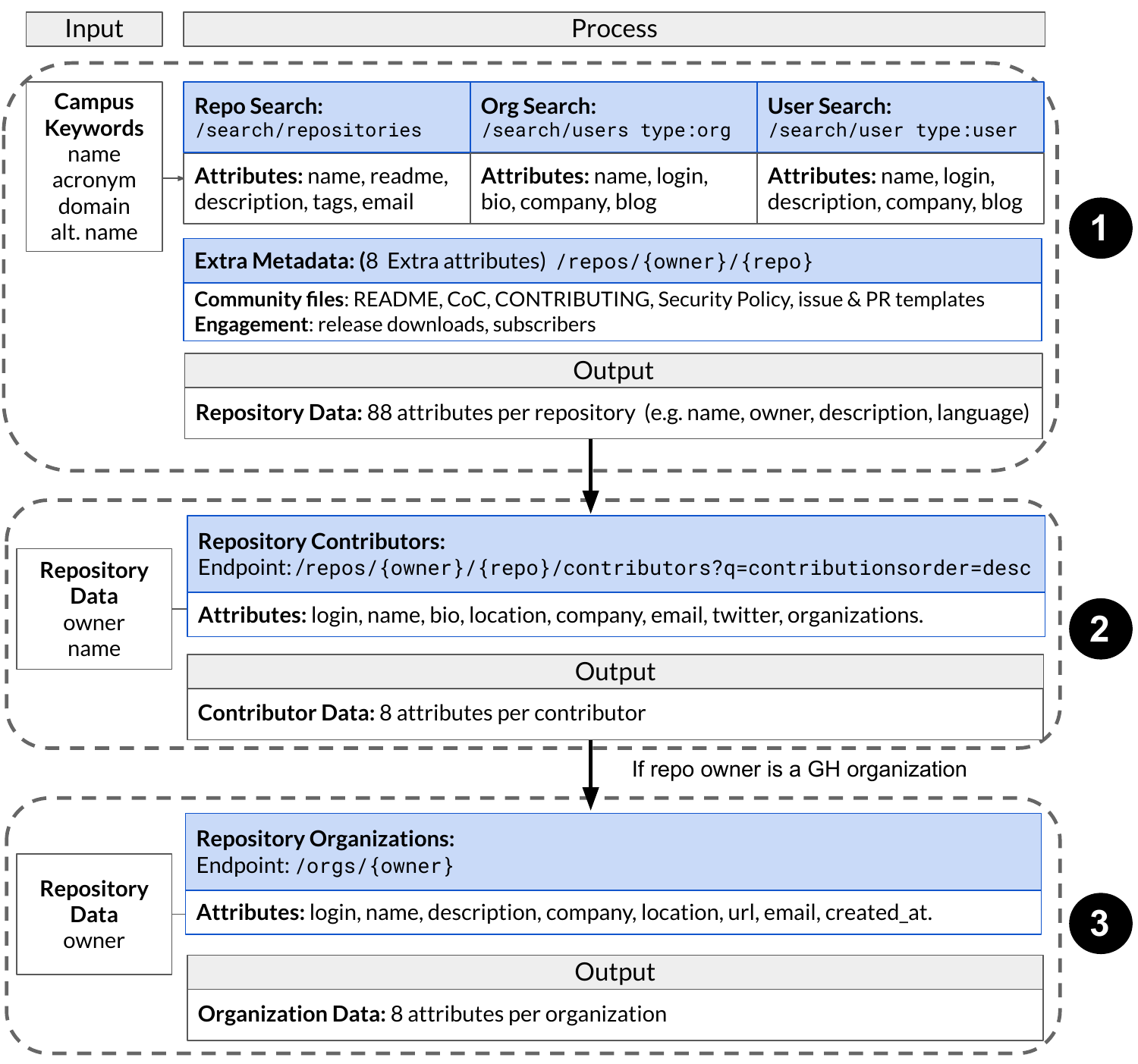}
  \caption{Data collection phases with GitHub's REST API.}
  \label{fig:scraping_diagram}
\end{figure}

To guide our analysis, we structure the data collection process into three distinct phases. Phase 1: Repository Data, where we collect core metadata and community files from GitHub repositories; Phase 2: Contributor Data, which enriches the dataset with information about individual contributors; and Phase 3: Organization Data which captures additional information from repository owners classified as organizations in GitHub. Figure~\ref{fig:scraping_diagram} provides an overview of this multi-phase workflow, highlighting the specific GitHub API endpoints and data elements involved in each stage.

\subsubsection{Phase 1: Repository Data}
 
The goal of this phase (illustrated in Figure~\ref{fig:scraping_diagram} phase \circled{1}) is to identify repositories that might be associated with an institution among those available on GitHub. Because this step deliberately favors recall over precision, we retain all matches and defer removing false positives to the filtering section.

To achieve this, we perform \textbf{three distinct but complementary searches} using GitHub’s REST API:
(1) a repository search,
(2) an organization search, and
(3) a user search.

\textbf{Repository search:} We begin by directly querying GitHub’s repository search endpoint (\texttt{/search/repositories}) using institution-specific lexical cues. For each University of California (UC) campus, we construct a set of keywords consisting of the university \texttt{name}, \texttt{acronym}, \texttt{email domain}, and a common \texttt{alternate name} (e.g., ``UC Berkeley''), as shown in Table~\ref{tab:uc-campuses}. These keywords are matched against five repository attributes: \texttt{name}, \texttt{description}, \texttt{README}, \texttt{tags}, and \texttt{email}. For the first four attributes, we query using all keywords, while for the \texttt{email} attribute we search only using the university domain. 

\textbf{Organization search:} In parallel, we query GitHub’s organization search endpoint (\texttt{/search/users} with \texttt{type:Organization}) to identify organizations whose profiles suggest institutional affiliation. Using the same set of campus specific keywords, we search across organization \texttt{name}, \texttt{login}, \texttt{description}, \texttt{company}, and \texttt{blog} fields. Repositories owned by the organizations returned by these queries are collected and added to the candidate repository set.

\textbf{User search:} Finally, we query the user search endpoint \\(\texttt{/search/users} with \texttt{type:User}) to identify individual developers whose profiles indicate a possible university affiliation. We search across user \texttt{name}, \texttt{login}, \texttt{bio}, \texttt{company}, and \texttt{blog} fields using the same institutional keywords. For each matched user, we collect all repositories owned by that user and include them in the dataset.

\begin{table}[htbp]
\centering
\scriptsize
\setlength{\tabcolsep}{3pt}
\resizebox{\columnwidth}{!}{%
\begin{tabular}{@{}l l l l@{}}
\toprule
\textbf{University Name} & \textbf{Acronym} & \textbf{Domain} & \textbf{Alt. Queries} \\
\midrule
University of California, Berkeley        & UCB  & berkeley.edu   & UC Berkeley \\
University of California, Davis           & UCD  & ucdavis.edu    & UC Davis \\
University of California, Irvine          & UCI  & uci.edu        & UC Irvine \\
University of California, Los Angeles     & UCLA & ucla.edu       & UC Los Angeles \\
University of California, Merced          & UCM  & ucmerced.edu   & UC Merced \\
University of California, Riverside       & UCR  & ucr.edu        & UC Riverside \\
University of California, Santa Barbara   & UCSB & ucsb.edu       & UC Santa Barbara \\
University of California, Santa Cruz      & UCSC & ucsc.edu       & UC Santa Cruz \\
University of California, San Diego       & UCSD & ucsd.edu       & UC San Diego \\
University of California, San Francisco   & UCSF & ucsf.edu       & UC San Francisco \\
\bottomrule
\end{tabular}%
}
\caption{Keyword queries.}
\label{tab:uc-campuses}
\end{table}

By combining repositories discovered directly through repository metadata with those obtained indirectly via organization and user ownership, this phase produces a high recall set of candidate repositories potentially associated with UC campuses.

The initial data scraping collects most of the metadata needed for our analysis (80 attributes in total), all based on the values provided by GitHub’s API. 
We supplement the initial search with additional API requests to collect specific repository features aligned with GitHub’s recommended community standards \cite{github-guidelines}. These include the presence of key community files, such as a README, code of conduct, contributing guide, security policy, issue and pull request templates, which we search using their most common filenames and paths \cite{github-community}. Lastly, we collect additional engagement indicators: release downloads and subscriber count. This data helps assess the project's maturity, community support, and open source practices.

\subsubsection{Phase 2: Contributor Data}

The second step is to collect information about the contributors to each repository. Having this information is essential for associating the projects with specific universities, since contributor profiles may contain affiliation data that validate a repository's connection to a specific campus. This helps also filter out some false positives, cases where there are keyword matches in the repository metadata but the repository does not necessarily seem affiliated with the institution. For example, repositories that reference or aggregate content from other (or multiple) universities. We collect the following attributes for each contributor (when available) from repositories identified in Phase 1: \texttt{username}, \texttt{name}, \texttt{bio}, \texttt{company} and \texttt{email}. The Contributor Data workflow is shown in Figure~\ref{fig:scraping_diagram}, phase~\circled{2}.

\subsubsection{Phase 3: Organization Data}

Finally, we collect additional information about the repository owner \textit{if} it is classified as an organization on GitHub (rather than an individual user).
Perhaps the clearest indicator of the affiliation of a repository with a university campus is when the organization that owns it (in GitHub) includes the university domain in its listed website or email address. We refer to these as high-confidence attributes, since they provide direct evidence of institutional affiliation.

We collect data from organizations using the GitHub user API endpoint \texttt{/users/owner}, retrieving the attributes: \texttt{login, name, company, description, email, url}. The organization data collection workflow is summarized in Figure~\ref{fig:scraping_diagram} phase \circled{3}.

\subsection{Data Filtering}
\label{subsec:data-filtering}

These three discovery phases yield a large volume of repositories, contributors, and organizations (on the order of hundreds of thousands) spanning all University of California campuses. This broad approach is designed for high recall, but inevitably includes many repositories that are not affiliated with the University of California. For instance, some repositories mention multiple universities or list academic programs, matching campus-related keywords despite lacking any apparent connection to the UC system. To prepare our dataset for analysis, we first clean the data to remove non informative repositories and then apply a filtering step to estimate the likelihood that each repository is genuinely associated with a university; the following subsections describe these steps in detail.

\subsubsection{Data Cleaning}

Our initial data cleaning step consists in removing repositories that are forks, archived, templates, or have zero reported size (repositories with no files), as these repositories do not represent original, active development and can introduce noise that affects repository counts and biases analyses of affiliation and activity. All of these attributes were collected during the initial data scraping process.


\subsubsection{Exploring Filtering Strategies}
Motivated by the broader evolution of classification methods from rule-based and feature-engineered approaches, to embedding-based supervised models, and more recently to large language models (LLMs) capable of contextual reasoning, we evaluated multiple filtering strategies with increasing ability to capture contextual signals. 

To compare these approaches we conducted all experiments using data from three UC campuses: UC Santa Cruz (UCSC), UC San Diego (UCSD), and UC Santa Barbara (UCSB). These campuses were selected to reflect institutional diversity within the UC system. Based on undergraduate enrollment sizes~\cite{uc_enrollment_2024}, UCSD represents a large campus, UCSB a medium-sized campus, and UCSC a smaller campus. Below, we summarize each approach and motivate our final choice. A more detailed description of the alternative methods is available in Appendix~\ref{app:filtering}.

\textbf{Score-Based Classifier (SBC):} As a baseline we implemented a rule-based scoring system inspired by prior work in feature engineering \cite{Munaiah2017}. This method assigns heuristic scores to matches between university related keywords and repository attributes (repository metadata, contributor profiles and organization information). While this approach is intuitive and easy to implement, it's incapable of distinguishing substantive institutional affiliation from superficial keyword matches leading to high false positive rate.

\textbf{Embedding-Based Machine Learning:} To incorportate semantic context beyond exact keyword matches, we explored a supervised learning approach using text embeddings that came from aggregated repository metadata. The embeddings were used as inputs for standard machine learning classifiers (neural networks, random forests, and support vector machines). Although the accuracy of the classification improved, the overhead that the manual labeling of the training set required motivated us to explore alternatives that did not require manual annotation. 

\textbf{Large Language Model Filtering:} Our final and most effective approach leverages large language models (LLMs) to asses repositories for university affiliation. We evaluated multiple OpenAI models: \texttt{gpt-3.5}, \texttt{gpt-4o}, and \texttt{gpt-5-mini}. While all three models were capable of reasoning over repository metadata, \texttt{gpt-5-mini} provided the best classification quality and cost-effectiveness.

\begin{figure*}[htb]
  \centering
  \includegraphics[width=0.8\textwidth]{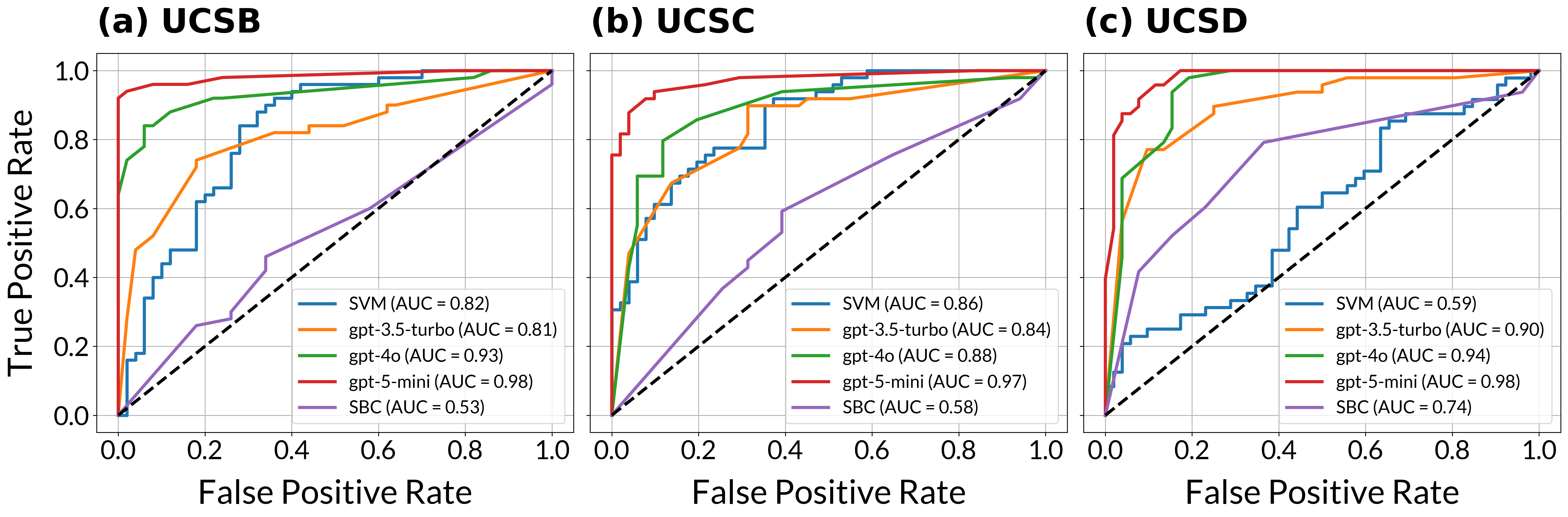}
  \caption{ROC curves for UCSB, UCSC, and UCSD for the five methods evaluated. \texttt{gpt-5-mini} achieves the highest area under the curve (AUC) for the three campuses, indicating strong classification performance.}
\label{fig:roc_all}
\end{figure*}

To evaluate the predictions of the different methods, we manually labeled 100 random data points from each of the 3 campuses. Figure~\ref{fig:roc_all} presents a ROC based comparison of these methods. A detailed cost and time comparison is provided in Appendix \ref{app:cost}

\subsubsection{Selected Filtering Method: gpt-5-mini}

Based on our evaluation of multiple filtering strategies, we adopt large language models (LLMs) as our primary method for determining repository university affiliation and filtering out false positives. We first formalize a definition of what it means for a repository to be affiliated with a university. Then, we use this definition to prompt the model via the OpenAI API with detailed contextual information about each repository, and relevant campus-related keywords.

\medskip
\noindent \textbf{(1) Defining "Affiliation":} Our first step is to define what it means for a repository to be \textbf{affiliated} with a university. Informed by our dataset, we identify a \textbf{repository’s signature}: Key characteristics that consistently indicate a repository's affiliation with a university, such as institutional ownership, contributor ties to the campus, and explicit mentions of university work or projects. This allows us to distinguish meaningful connections from incidental keyword matches.

\begin{adjustwidth}{.1cm}{.1cm}
\begin{smalldefn}
\label{def1}
A repository is considered affiliated with a university if there is clear, public evidence that its development, maintenance, or governance is connected to the university. This includes any of the following criteria:

\textbf{1. Research or Academic Unit Development}:  
Developed or maintained by a research group, lab, center, academic department, or institute that is formally part of the university.  
\emph{Evidence} includes README statements such as ``developed at \dots'', links to lab websites, or institutional pages under the university domain.

\textbf{2. Contributor Affiliation (Explicit Evidence)}:  
One or more key contributors (maintainers, lead developers, or primary committers) are affiliated with the university.  
\emph{Evidence} includes: (1) university email addresses, (2) public profiles or bios naming the university, or (3) README sections listing team members and their university affiliation.

\textbf{3. Institutional or Administrative Development}:  
Developed or maintained by an official, non-academic university unit (e.g., libraries, Open Source Program Offices, IT departments, or research computing groups).

\textbf{4. Official Ownership, Sponsorship, or Endorsement}:  
Owned, sponsored, or explicitly endorsed by the university.  
\emph{Evidence} includes: (1) hosting under an official university GitHub organization, (2) README or website statements indicating university sponsorship or ownership, or (3) copyright notices naming the university.

\textbf{5. Documented Collaboration or Partnership}:  
The README or project website explicitly states a collaboration, partnership, or joint development effort with the university or one of its units.

\textbf{6. University-Linked Project Infrastructure}:  
The project’s official homepage, documentation site, or primary web presence is hosted under the university’s domain or a clearly associated subdomain.

\textbf{7. Educational Outreach and Online Courses}:  
Online learning initiatives affiliated with the university, including: (1) repositories linked to online specializations or courses offered on platforms such as Coursera or edX, and (2) course materials, code examples, or tools developed specifically for such offerings.
\end{smalldefn}
\end{adjustwidth} 
\medskip
\noindent \textbf{(2) Building the prompt: } The prompt asks the model to return a probability value between 0 and 1 indicating the likelihood that the repository is affiliated with the university, along with a brief explanation justifying its assessment. To construct the prompt, we aggregate multiple metadata components: repository metadata \( \text{repo}_i \), organizational metadata \( \text{org}_i \), and information about the top contributors \( \text{contrib}_i^{(1)} \) and \( \text{contrib}_i^{(2)} \). Due to the model’s input token limit, each text field is truncated to a maximum of 20000 characters, retaining the first 15000 characters and the final 5000 characters. This truncation primarily affects README files. Based on manual examination of a sample of repositories, we observed that acknowledgements, credits, and contact information that indicate institutional affiliation frequently appear near the end of README files. An example of the constructed prompt is as follows:
\begin{tcolorbox}[
  colback=gray!10,    
  colframe=gray!60,   
  boxrule=0.5pt,      
  arc=3pt,            
  left=6pt,           
  right=6pt,          
  top=6pt,            
  bottom=6pt,         
  enhanced,
  breakable 
]
\begin{adjustwidth}{1em}{1em}
\scriptsize
\ttfamily
You are tasked with determining the likelihood that a GitHub repository belongs to a university based on the following definition:
\begin{itemize}
    \item Definition: <Definition 1>
    \item Here is the information about the repository: <repository information>
    \item And here is the university context: <university information> 
\end{itemize}

\noindent 
Based on this information and the definition above: 
\begin{itemize}
    \item Estimate the probability between 0 and 1 (e.g., 0.87) representing how likely it is that the repository belongs to the university.
    \item Provide a brief explanation (1-2 sentences) justifying your answer.
\end{itemize}
Your response must be formatted exactly like this:
\begin{itemize}
    \item \texttt{Probability: <value between 0 and 1>}
\item \texttt{Explanation: <your explanation here>}
\end{itemize}

\end{adjustwidth}
\end{tcolorbox}


\subsection{Data Categorization}

After identifying repositories affiliated with a university, we further classify them by \textbf{project type} to better understand the purpose of institutional open source contributions. This classification enables more granular insight into how universities engage with open source, whether through tool development, education, web hosting, data set creation, or documentation. We designed a set of categories inspired by prior work on repository classification~\cite{classifyhub}, with several adjustments to improve compatibility with LLMs, including the addition of concrete examples and keyword indicators for each category to provide a clearer context and reduce ambiguity.

\begin{adjustwidth}{.1cm}{.1cm}
\begin{smalldefn}
\label{def:repo-categories}
You must classify each GitHub repository into exactly one of the following categories based on its primary purpose.
When multiple purposes are present follow the precedence rules described below.

\textbf{DEV}: A repository primarily used for the development and maintenance of a software artifact, including tools,
libraries, components, applications, services, or APIs. The presence of documentation, a website, or example code
does not override this classification if active software development is the main function.

\textbf{EDU}: A repository primarily used for educational purposes, including course-related materials (e.g.,
assignments, labs, class projects, or course websites), teaching infrastructure, or instructional content explicitly
tied to a course, workshop, or training program. This category includes teaching demos and repositories created to
support internships, tutorials, or learning exercises, provided their main goal is instruction rather than
production use. If a repository contains software or an application developed as part of completing a course or
academic requirement, it is classified as EDU, even if it resembles a standalone or production-style software project.

\textbf{DOCS}: A repository primarily used to store or track non-educational documents, such as reports, policies,
white papers, specifications, or meeting notes. 

\textbf{WEB}: A repository primarily used to host a public-facing website or informational page, such as a project
homepage, documentation website, or static or CMS-based informational site. This includes repositories built with
static site generators (e.g., Jekyll or Hugo), as well as site-specific styles, layouts, or templates, when the main
purpose is presentation rather than software development. Reusable themes or design systems intended for general
adoption are excluded and classified as DEV. Personal websites, portfolios, or ``about me'' pages are classified as
OTHER.

\textbf{DATA}: A repository primarily used to store, curate, or distribute datasets, including research datasets,
benchmarks, or data collections (including images). Lightweight scripts for data loading or inspection do not change
this classification.

\textbf{OTHER}: A repository that does not clearly align with any of the above categories, such as empty repositories,
personal experiments, configuration-only repositories, or miscellaneous content without a clear primary purpose.
\end{smalldefn}
\end{adjustwidth}


The classification prompt provides the LLM with a standardized set of definitions and relevant repository metadata: full name, description, README content, number of stars, forks, and contributors. The model is then asked to output the best fit category along with a brief explanation. The prompt used is shown below:
\begin{tcolorbox}[
  colback=gray!10,    
  colframe=gray!60,   
  boxrule=0.5pt,      
  arc=3pt,            
  left=6pt,           
  right=6pt,          
  top=6pt,            
  bottom=6pt,         
  enhanced,
  breakable 
]
\begin{adjustwidth}{1em}{1em}
\scriptsize
\ttfamily
You must classify each GitHub repository into exactly one of the following categories: <Definition 2>

Choose the most appropriate category based on the repository information.

\noindent
\begin{itemize}
    \item Here is the information about the repository: <repository information>

    \item Your task: Return only the predicted category (one of: DEV, EDU, DOCS, WEB, DATA, OTHER), and a short explanation.
    \item Your response must be formatted exactly as:
    \begin{itemize}
        \item \texttt{Category: <one of the 6 categories>}  
        \item \texttt{Explanation: <brief explanation>}
    \end{itemize}
\end{itemize}
\end{adjustwidth}
\end{tcolorbox}
\section{Results}

\subsection{Discovery}

Table~\ref{tab:repo-stats} illustrates the breakdown of the repositories collected on the ten campuses of the University of California (UC). We found 236037 unique repositories with our preliminary collection, and given the scale, manual verification and filtering of each repository is impractical. 

\begin{table}[h]
  \centering
  \footnotesize
  \begin{tabular}{lrrrrr}
    \toprule
    \textbf{University} & \textbf{Repos} & \textbf{Contributors} & \textbf{Organizations} \\
    \midrule
    UCB & 36552 & 34981 & 1108 \\
    UCD & 26549 & 16400 & 695 \\
    UCI & 28730 & 20748 & 657 \\
    UCLA & 24917 & 18397 & 651 \\
    UCM & 10132 & 9405 & 392 \\
    UCR & 16050 & 10492 & 432 \\
    UCSB & 19907 & 12051 & 651 \\
    UCSC & 24181 & 14377 & 662 \\
    UCSD & 42729 & 24981 & 908 \\
    UCSF & 6290 & 4731 & 373 \\
    \midrule
    \textbf{Total} & 236037 & 166563 & 6529  \\
    \bottomrule
        \\
  \end{tabular}
  \caption{Repository and contributor statistics by UC campus.}
  \label{tab:repo-stats}
\end{table}

\subsection{Filtering}
We now present the results of the filtering procedure for three universities with available test data (UCSD, UCSC, and UCSB) evaluating the pipeline’s ability to correctly identify institutionally affiliated repositories and filter out false positives.

\subsubsection{Accuracy Analysis}
As discussed in Section~\ref{subsec:data-filtering}, \texttt{gpt-5-mini} achieved the best performance at filtering out false positives. We determine the optimal classification threshold for this model by maximizing Youden's J statistic \cite{YoudensJ}, and we evaluate its performance using standard metrics, including accuracy and the F1 score. The results are summarized in Table~\ref{tab:filtering-performance}.

\begin{table}[h]
\centering
\footnotesize
\setlength{\tabcolsep}{5pt}
\resizebox{\columnwidth}{!}{%
\begin{tabular}{lcccc}
\toprule
\textbf{University} 
& \textbf{F1-score} 
& \textbf{Accuracy} 
& \textbf{False Pos.} 
& \textbf{False Neg.} \\
& & & \textbf{(\%)} & \textbf{(\%)} \\
\midrule
UCSB & 0.90 & 0.91 & 0\% & 9\% \\
UCSC & 0.90 & 0.91 & 1\% & 8\% \\
UCSD & 0.93 & 0.94 & 2\% & 4\% \\
\bottomrule
\end{tabular}
}
\caption{Filtering performance across three UC campuses for \texttt{gpt-5-mini}.}
\label{tab:filtering-performance}
\end{table}

As shown in Table~\ref{tab:filtering-performance}, both accuracy and F1 score exceed 90\% across all three campuses, indicating that \texttt{gpt-5-mini} performs well at identifying university-affiliated repositories. The majority of classification errors are false negatives, repositories that should be marked as affiliated but are missed by the language model. Representative examples include R\citeR{openfermion}, R\citeR{openfermion-pyscf}, R\citeR{design-patterns-julia}, R\citeR{c-specialization}, and R\citeR{vsdsram}. In many of these cases, the university connection is subtle, such as a faculty member being mentioned only at the end of a README file or brief, single-line acknowledgments that require careful attention. The language model appears to struggle particularly when repositories have contributors from multiple institutions, making the institutional affiliation less explicit. These observations suggest that when affiliation evidence is difficult to locate, language models are more likely to fail to identify it.

A smaller number of false positives were also observed. For example, the repository R\citeR{donoyo} contains multiple links using a university domain and hosts scripts and data; however, it lacks a clear or substantive connection to the university itself. 

\subsubsection{Affiliated Repositories Per Campus}

We use the \texttt{gpt-5-mini} model to filter false positives from the repositories of the remaining seven campuses, yielding the results shown in Table~\ref{tab:predictions}. The table presents the total number of repositories, the number remaining after data cleaning, and the number of repositories classified as affiliated.

\begin{table}[htb]
\centering
\footnotesize
\setlength{\tabcolsep}{3pt} 
\begin{tabular}{lrrrr}
\toprule
\textbf{University} & \textbf{Repos} & \textbf{Cleaned} & \textbf{\# Affil.} & \textbf{\% Affil.} \\
\midrule
UCB  & 36552 & 25810 & 12232 & 47.4\% \\
UCD  & 26549 & 20526 & 8736  & 42.6\% \\
UCI  & 28730 & 21958 & 8134  & 37.0\% \\
UCLA & 24917 & 18088 & 9859  & 54.5\% \\
UCM  & 10132 & 8395  & 1989  & 23.7\% \\
UCR  & 16050 & 12603 & 5672  & 45.0\% \\
UCSB & 19907 & 14679 & 7926  & 54.0\% \\
UCSC & 24181 & 18266 & 6264  & 34.3\% \\
UCSD & 42729 & 28416 & 18163 & 63.9\% \\
UCSF & 6290  & 4537  & 2665  & 58.7\% \\
\midrule
\textbf{Total} & \textbf{236037} & \textbf{173278} & \textbf{81640} & \textbf{47.1\%} \\
\bottomrule
\end{tabular}
\caption{Predicted affiliated repositories per university using \texttt{gpt-5-mini}.}
\label{tab:predictions}
\end{table}

Overall, only about half of the cleaned repositories are predicted to be truly affiliated with a UC campus, highlighting the substantial amount of noise present in the initial dataset. A significant portion of projects, approximately 27\% of the original 236K repositories were excluded during the cleaning stage because they corresponded to templates, forks, archived repositories, or empty projects, reflecting the need for careful preprocessing when working with this large scale repository data. Affiliation rates vary considerably across campuses: while UCSD, UCLA, UCSB, and UCSF show affiliation rates above 50\%, UCM exhibits a notably lower rate (23.7\%). This lower affiliation rate is likely influenced by the relatively small size of the campus. As one of the smallest institutions in the UC system, UC Merced has less research contributions, which results in fewer affiliated open source projects overall. San Diego (UCSD) stands out as the campus with the largest absolute number of affiliated repositories, which is consistent with its scale and research intensity as one of the largest research universities in the system. 

\subsection{Project Type Classification}
We now present the results of the project type categorization, which identifies the main purpose of each affiliated repository using predefined category labels. 

\subsubsection{Accuracy Analysis}

Given the overall better performance of \texttt{gpt-5-mini} in the previous section, we select this model for project type classification. We first evaluate its performance on three campuses (UCSB, UCSC, and UCSD), for which we manually label 100 repositories per campus, sampled from the set of repositories previously predicted as affiliated. These manual labels are used to assess the accuracy of the GPT-based predictions. We do not balance the sample by project type; instead, we preserve the natural distribution of project types to evaluate whether the model’s predicted class distribution aligns with the underlying distribution in the data.

We report results using accuracy per project type, calculated against the manually labeled test set. Table~\ref{tab:type-classification-results} shows the model's performance across the three universities, along with the number of repositories labeled for each project type in the test subsets.

\begin{table}[h]
  \centering
  \footnotesize
  \setlength{\tabcolsep}{3pt} 
  \resizebox{\columnwidth}{!}{%
  \begin{tabular}{lcc|cc|cc}
    \toprule
    \textbf{Project Type} 
    & \multicolumn{2}{c|}{\textbf{UCSB}} 
    & \multicolumn{2}{c|}{\textbf{UCSC}} 
    & \multicolumn{2}{c}{\textbf{UCSD}} \\
    \cmidrule(lr){2-3} \cmidrule(lr){4-5} \cmidrule(lr){6-7}
     & \textbf{Count} & \textbf{Acc.} 
     & \textbf{Count} & \textbf{Acc.} 
     & \textbf{Count} & \textbf{Acc.} \\
    \midrule
    DATA   & 8   & 0.62 & 0   & 0.00 & 4   & 0.50 \\
    DEV    & 26  & 0.92 & 30  & 0.90 & 21  & 0.86 \\
    DOCS   & 0   & 0.00 & 3   & 0.33 & 1   & 0.00 \\
    EDU    & 51  & 0.98 & 50  & 0.96 & 65  & 0.98 \\
    OTHER  & 12  & 0.92 & 9   & 0.89 & 6   & 0.83 \\
    WEB    & 3   & 0.67 & 8   & 0.75 & 3   & 1.00 \\
    \midrule
    \textbf{Total} & 100 & 0.92 & 100 & 0.90 & 100 & 0.92 \\
    \bottomrule
  \end{tabular}%
  }
  \caption{Project type classification results per university. Each cell shows the number of repos and accuracy.}
  \label{tab:type-classification-results}
\end{table}

 Overall, \texttt{gpt-5-mini} achieves consistently high accuracy across the three universities (90–92\%), comparable to the performance observed for affiliation prediction, suggesting that the model generalizes well to this downstream classification task. The highest accuracies are observed for \texttt{EDU} and \texttt{DEV} repositories, which is expected given the presence of strong and explicit indicators of purpose in these projects, such as course identifiers, assignment structures, or deployment instructions. In contrast, \texttt{DATA} and \texttt{WEB} repositories are more challenging to classify accurately without explicit metadata about the data types within the repositories, and their README files or descriptions can be ambiguous, sometimes resembling documentation or development repositories. \texttt{DOCS} repositories are relatively rare in the test sets and exhibit lower accuracy, reflecting their inherent ambiguity and overlap with \texttt{WEB} and \texttt{DEV} categories, particularly when documentation is colocated with code or hosted as a static site. 

\subsubsection{Distribution of Repositories by Project Type}
We use \texttt{gpt-5-mini} to classify the affiliated repositories for all 10 universities and analyze their distribution across the predefined repository types for each UC campus, as summarized in Table~\ref{tab:project-type-distribution}.

\begin{table*}[t]
  \centering
  \footnotesize
  \setlength{\tabcolsep}{4pt}
  \begin{tabular}{lrrrrrrr}
    \toprule
    \textbf{University} 
    & \textbf{DEV} 
    & \textbf{EDU} 
    & \textbf{DOCS} 
    & \textbf{WEB} 
    & \textbf{DATA} 
    & \textbf{OTHER} 
    & \textbf{Total} \\
    \midrule
    UCB  & 47.0\% & 38.7\% & 0.9\% & 3.2\% & 2.1\% & 8.1\% & 12232 \\
    UCD  & 44.0\% & 39.5\% & 1.2\% & 3.1\% & 3.1\% & 9.1\% & 8736 \\
    UCI  & 49.7\% & 33.5\% & 0.7\% & 3.0\% & 1.6\% & 11.4\% & 8134 \\
    UCLA & 48.2\% & 36.4\% & 1.2\% & 3.6\% & 2.5\% & 8.2\% & 9859 \\
    UCM  & 38.3\% & 45.3\% & 0.8\% & 3.4\% & 3.2\% & 8.9\% & 1989 \\
    UCR  & 37.4\% & 46.6\% & 0.5\% & 4.0\% & 1.7\% & 9.9\% & 5672 \\
    UCSB & 27.6\% & 60.3\% & 0.8\% & 2.9\% & 2.5\% & 6.0\% & 7926 \\
    UCSC & 42.3\% & 43.3\% & 0.8\% & 2.8\% & 1.9\% & 8.9\% & 6264 \\
    UCSD & 39.9\% & 46.0\% & 0.8\% & 2.9\% & 2.0\% & 8.4\% & 18163 \\
    UCSF & 63.0\% & 18.3\% & 1.9\% & 4.8\% & 4.5\% & 7.5\% & 2665 \\
    \midrule
    \textbf{TOTAL} & 42.9\% & 42.1\% & 0.9\% & 3.2\% & 2.3\% & 8.6\% & 81640 \\
    \bottomrule
  \end{tabular}
  \caption{Percentage distribution of project types by university, with total affiliated repository counts.}
  \label{tab:project-type-distribution}
\end{table*}

 Overall, the distribution is dominated by \texttt{DEV} and \texttt{EDU} repositories, which together account for approximately 85\% of all affiliated projects (42.9\% DEV and 42.1\% EDU). This balance suggests that GitHub serves both as a platform for active software development and as a central infrastructure for instructional use across the UC system. Development repositories are particularly prevalent at research-intensive campuses such as UCB and UCLA, where nearly half of affiliated projects fall into the \texttt{DEV} category.

Educational repositories constitute a substantial share at most campuses, with particularly high concentrations at UCSB (60.3\%), UCR (46.6\%), and UCM (45.3\%), reflecting the strong presence of course related materials, assignments, and student projects. In contrast, UCSF exhibits a different distribution: 63.0\% of its affiliated repositories are classified as \texttt{DEV}, while only 18.3\% fall into the \texttt{EDU} category. This pattern is consistent with UCSF’s role as a graduate only focused on health sciences. Since UCSF does not offer undergraduate degrees, it has a smaller
pool of students enrolled in coursework, leading to fewer educa-
tional repositories and a stronger emphasis on research-oriented
projects. Across all campuses, \texttt{WEB} and \texttt{DATA} repositories each represent a relatively small fraction of affiliated projects (typically 2–5\%), while \texttt{DOCS} repositories are rare, consistently accounting for around 1\% or less. 

\section{Insights on Affiliated Repositories}

With these predictions, we can gain insight into the types of practices and trends of university-affiliated repositories. 
We start by analyzing the programming languages and software licenses used in the affiliated repositories. 

\subsection{Programming Language Distribution}

\begin{figure}[htb]
    \centering
    \includegraphics[width=0.45\textwidth]{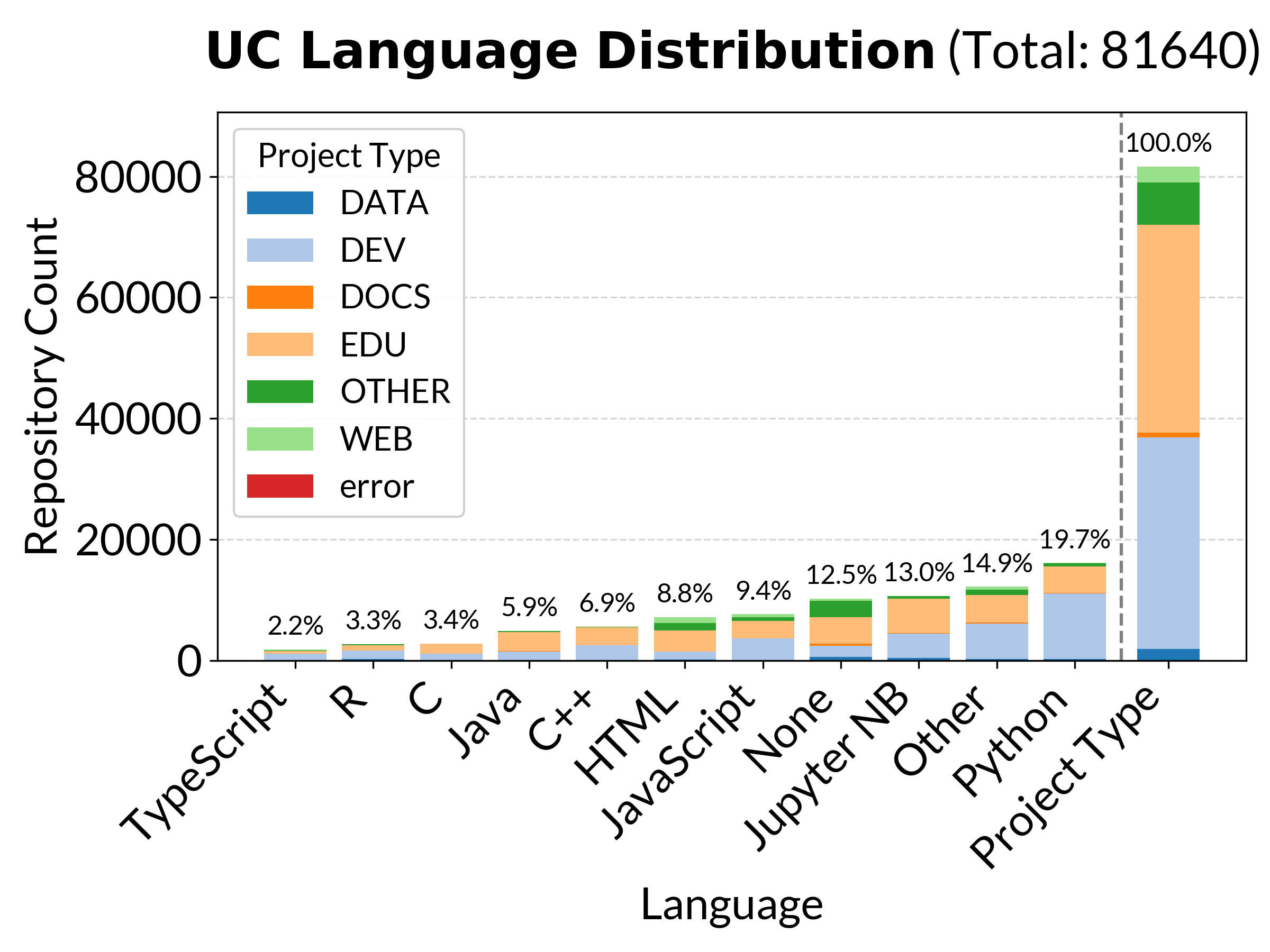}
    \caption{Programming language distribution across all ten University of California campuses. Languages representing less than 2\% of total usage were aggregated under “Other.” The bar on the right labeled “Project type” shows the overall distribution of repository types and serves as a baseline for interpreting language usage within each category.}

    \label{fig:language-distribution}
\end{figure}

Figure~\ref{fig:language-distribution} shows the aggregated distribution of programming languages in all affiliated repositories from the 10 UC campuses, broken down by project type.

 Python stands out as the language most commonly used in general, aligning with GitHub's 2024 Octoverse report \cite{octoverse2024}, which also ranked Python as the top language in public repositories. Although Python is the most dominant language within the development (DEV) repositories, it still represents only 17. 4\% of all repositories, reflecting the general diversity in language usage. Jupyter Notebook is also highly prevalent, especially in educational (EDU) repositories, where it serves as a popular format for teaching and interactive coursework. HTML is the most frequently used language in web-related (WEB) repositories, closely followed by JavaScript. Other notable languages include Java, mainly used in educational projects, C++, C, and R which appear across both DEV and EDU types.

\subsection{License Distribution}

\begin{figure}[htb]
    \centering
    \includegraphics[width=0.45\textwidth]{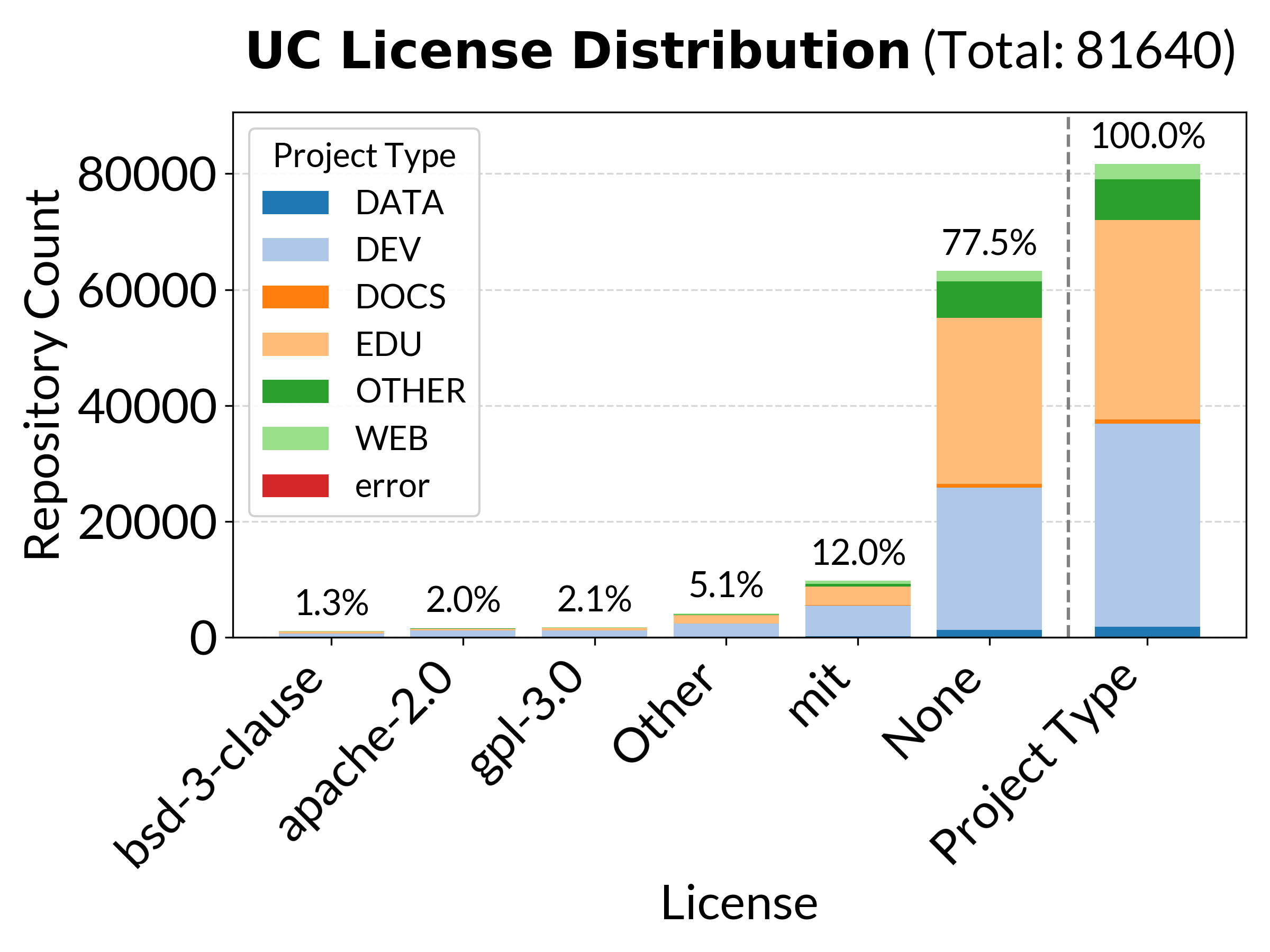}
    \caption{Software License usage across all 10 UCs. Licenses representing less than 1\% of total usage were aggregated under “Other.”  The bar on the right labeled “Project type” shows the overall distribution of repository types and serves as a baseline for interpreting distribution within each type.}
    \label{fig:license-distribution}
\end{figure}

Beyond language usage, understanding the licensing choices of repositories provides insight into how openly these projects are intended to be shared and reused. Figure~\ref{fig:license-distribution} shows the distribution of software licenses from the 10 UCs. A notable trend is that over 70\% of repositories appear to lack an explicit license. It is important to note that “no license” does not necessarily mean a repository has no license; in some cases, the GitHub API is simply unable to retrieve it. Many of these unlicensed repositories are development (DEV) projects, likely smaller research efforts or student works not intended for broad open source distribution. Almost 100\% of the repositories in the category 'OTHER' have no license, consistent with their varied and often informal content. Among repositories that specify a license, MIT is by far the most popular, covering 12\% of all repositories, representing more than one third of those that have any license, and is prevalent across both educational (EDU) and development projects (DEV), as well as a smaller portion of web projects. There is also notable license diversity beyond MIT, including GPL, Apache, and BSD licenses. At least 5\% use a variety of other licenses, each representing less than 1\% individually, suggesting a wide range of license choices across the dataset.

\subsection{Adoption of GitHub Community Standards}

\begin{figure}[htb]
    \centering
    \includegraphics[width=0.45\textwidth]{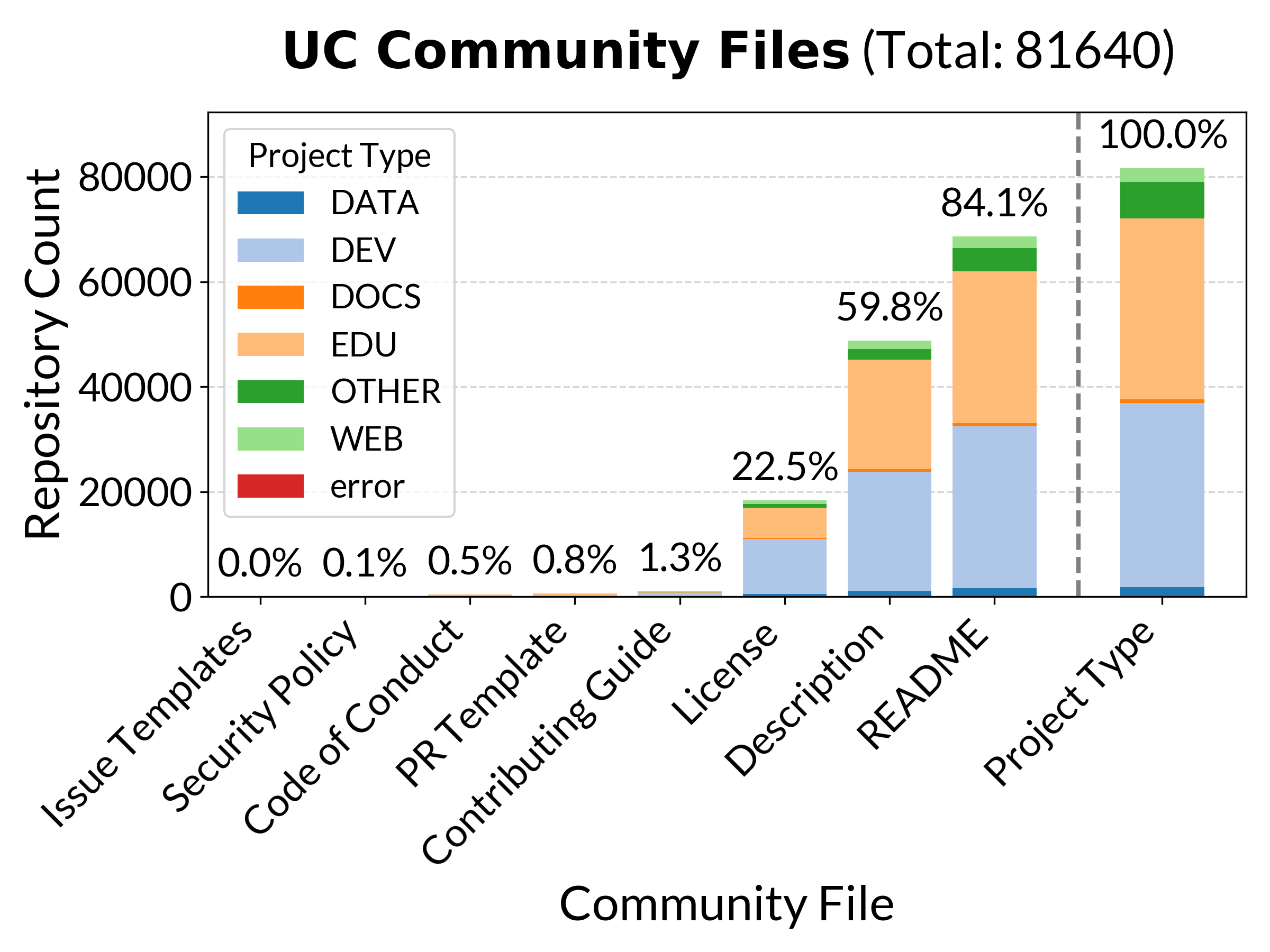}
    \caption{Presence of community files across all ten University of California campuses. The bar on the right labeled “Project type” shows the overall distribution of repository types and serves as a baseline for interpreting community file usage within each category.}
    \label{fig:features-distribution}
\end{figure}

We also aim to understand how people developing software within academic contexts engage with community practices. To do so, we analyze the presence of the community standards recommended by GitHub as indicators of openness, collaboration, and project sustainability. 

Figure~\ref{fig:features-distribution} illustrates the percentage of repositories that include GitHub’s recommended community standards~\cite{github-guidelines}: Description, README, License, Code of Conduct, Contributing Guide, Security Policy, Issue Templates, and Pull Request Templates.  Including these files is considered a best practice in open source development, as they help make a project easier to understand, use, and contribute.

Most repositories include a README file (around 85\%) and a description (around 60\%), making these the most commonly adopted community practices. Only about 23\% per university have an explicit license, but when looking at the other community guidelines, the percentages reduce considerably. Less than 2\% of the repositories provide contributing guidelines, codes of conduct, security policies or templates for issues and pull requests. This trend suggests that most of the repositories are small scale efforts, rather than mature open source tools that expect outside contributions. It may also reflect a general lack of awareness among students and researchers about open source best practices, even though they are essential to enabling collaboration and long term sustainability.

\begin{figure}[htb]
    \centering
    \includegraphics[width=.45\textwidth]{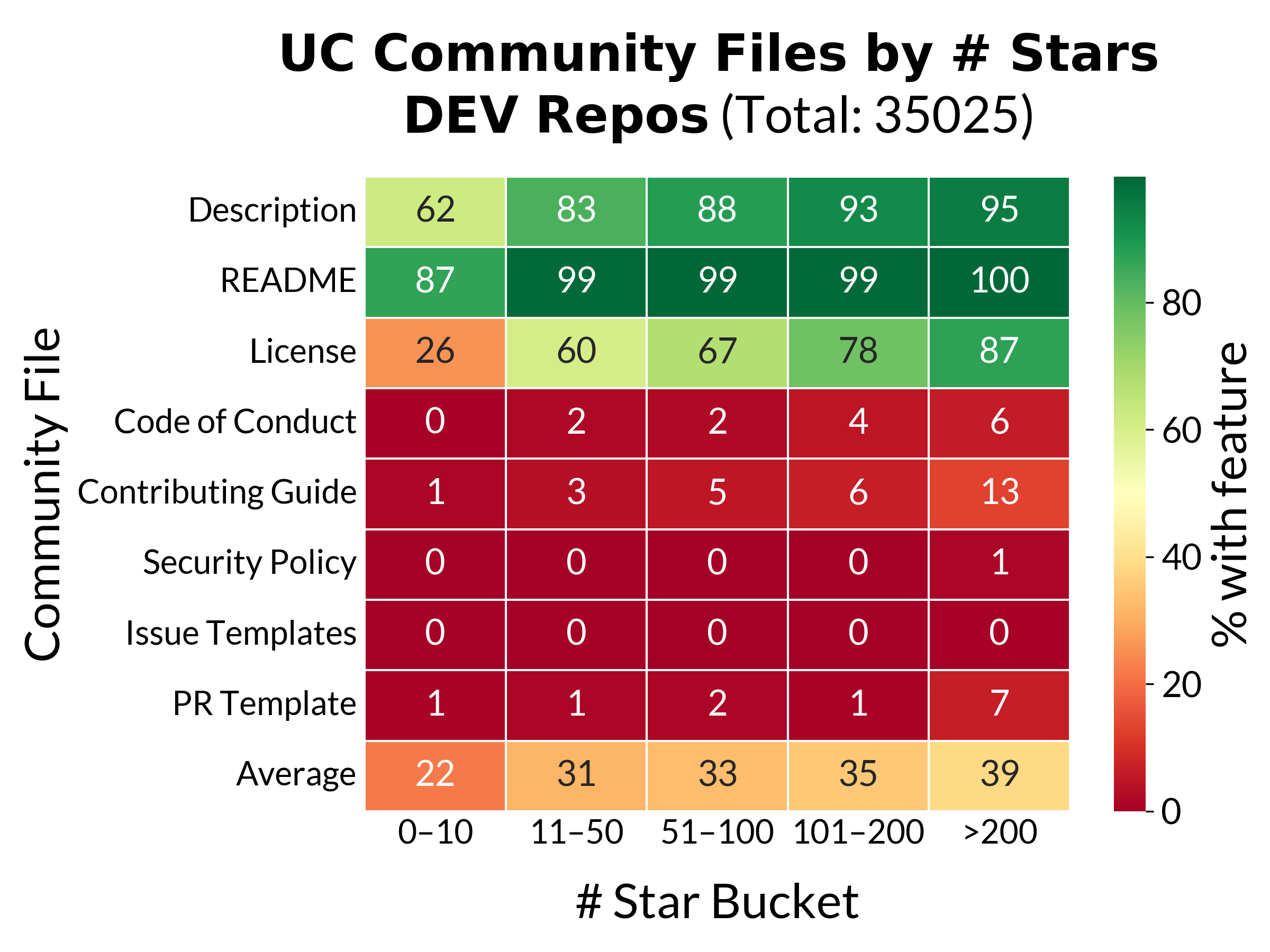}
    \caption{Presence of community files across all 10 UC campuses of DEV repositories by number of stars}
    \label{fig:heatmap-stars}
\end{figure}

We are particularly interested in the repositories classified as development (DEV), as these are the most likely to be intended for collaborative software development. Additionally, we want to understand how project popularity influences the adoption of community standards under the assumption that more popular projects may attract more contributors, and therefore require better documentation and engagement practices. We use the number of stars as a measure of popularity since they provide a useful indication of interest and community engagement, and we focus our analysis on DEV repositories across all UC campuses. Figure~\ref{fig:heatmap-stars} presents a heatmap showing the presence of community files in the DEV repositories, segmented by popularity. We group repositories into five star buckets based on their number of GitHub stars and compute the percentage of repositories in each bucket that include each community file.

As shown in the heatmap, most DEV repositories include a description and README file; however, these practices become nearly universal among more popular projects. Specifically, more than 90\% of repositories with more than 100 stars include a description and a README. Licensing shows a particularly strong correlation with popularity: almost 80\% of repositories with 50–100 stars declare a license, and this increases to 87\% for those with more than 200 stars. A similar upward trend is observed for the presence of a code of conduct and contributing guide, both of which increase monotonically across star buckets. However, their overall adoption remains limited: only 6\% of the projects include a code of conduct and 14\% include a contributing guide, even in the highest popularity tier. For pull request templates, the trend is less evident, with only a big rise in adoption once projects exceed 200 stars. Finally, security policies and issue templates remain rare across the board.

To further explore the relationship between project popularity and the adoption of community practices, we compute the average presence of community files across DEV repositories, grouped by star count. For this, we divide repositories into 20 evenly spaced buckets based on their number of stars, capping the maximum at 1000 to ensure a balanced scale (the most starred project in our dataset has over 40000 stars). For each bucket, we calculate the percentage of repositories that include each community file and then average these percentages across all community file types. 

\begin{figure}[htb]
    \centering
    \includegraphics[width=.45\textwidth]{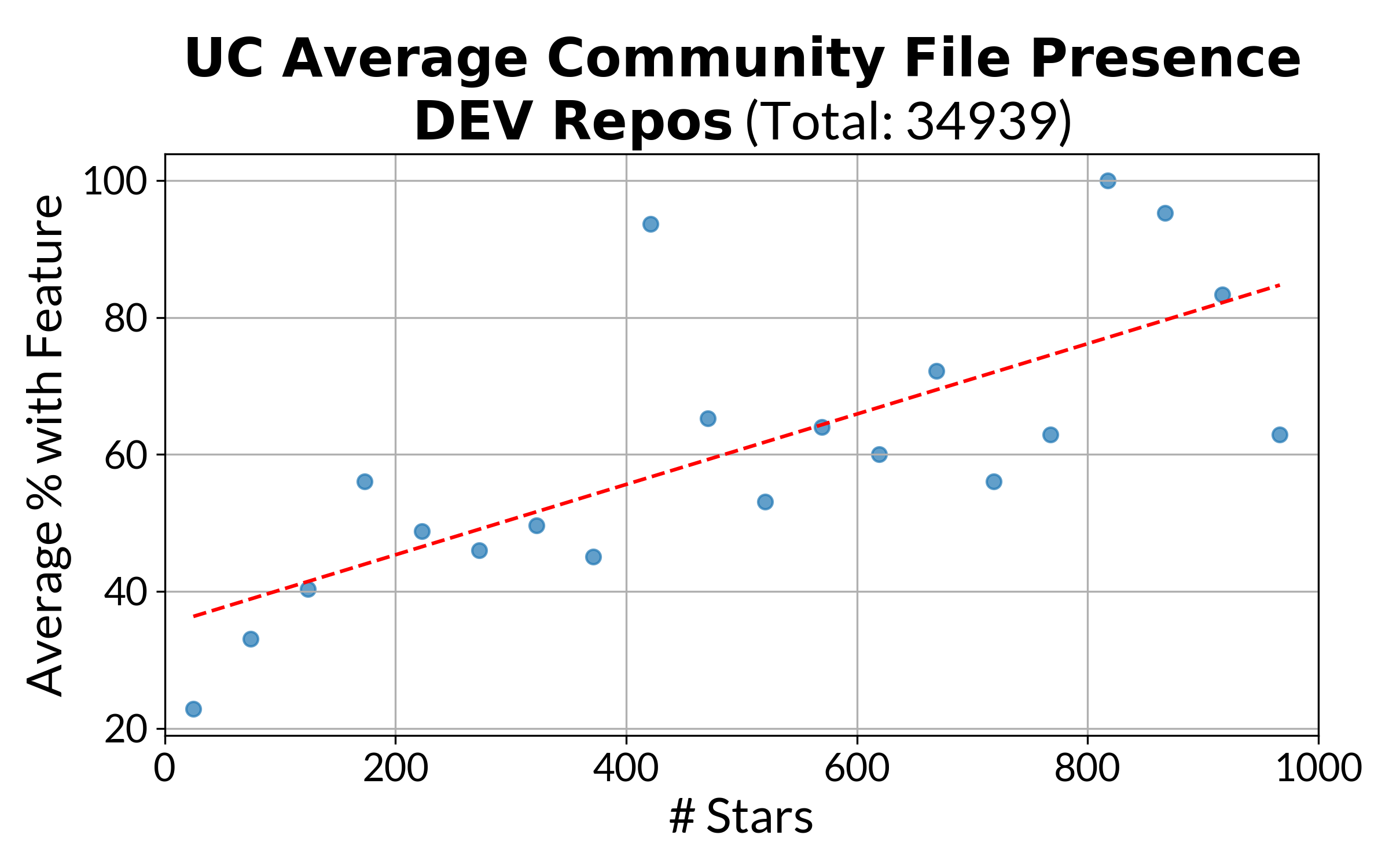}
    \caption{Average presence of community files in DEV repositories across all 10 UC campuses, grouped by GitHub star count (capped at 1000 stars).}
    \label{fig:scatterplot}
\end{figure}

The resulting scatter plot (Figure~\ref{fig:scatterplot}) shows a clear positive correlation: As the number of stars increases, so does the average presence of community files. 
Despite this correlation, the overall adoption of community standards remains relatively low. There are projects with more than 200 stars showing less than 50\% adoption, highlighting that many best practices of collaborative development are still underutilized throughout the UC ecosystem. There is a clear gap and an opportunity for improvement: increasing awareness and helping project maintainers navigate these practices could strengthen the open source culture within academic settings.

\section{Conclusions and Future Work}
In this work, we introduced a framework for identifying, filtering, and analyzing open source repositories affiliated with decentralized institutions. Our results reveal that many repositories initially retrieved are not truly affiliated with the studied institutions, and among those that are, most are educational (EDU) rather than development (DEV) projects. This distribution underscores both the diversity of institutional repositories and the necessity of automated filtering methods to enable meaningful and scalable analysis.

Our study allowed us to obtain a comprehensive picture of the open source landscape of the UC system. This analysis provides insights into community practices across affiliated repositories, identifying key gaps and opportunities where institutions, such as the Open Source Program Offices (OSPOs) can provide additional support and resources to strengthen these communities.

Although our case study focuses on the University of California (UC) system, the proposed pipeline is not specific to any single university system and can be applied to a wide range of institutions. Researchers can easily adapt it to different institutions by modifying a single configuration file containing institution specific keywords (e.g., names, acronyms, and domains). Once configured, the pipeline automatically collects, classifies, and analyzes affiliated repositories. By openly sharing our code, we aim to enable other institutions to map and better understand their open-source ecosystems.

As a next step, we will build a public browser of open source projects across the UC university systems, enabled by the fact that our database allows us to systematically identify high impact projects using engagement and popularity indicators such as stars, forks, and contributor activity. This capability makes it possible to move beyond discovery and toward prioritizing projects that are actively used and maintained.

This centralized browser will showcase the contributions of universities to open source scientific projects while also helping to surface missing or inconsistent practices and supporting maintainers in strengthening them. Notably, many institutions have already attempted to track and promote impactful open-source projects by building project registries but this work is typically performed manually and does not scale. Our approach provides an automated, extensible alternative that enables institutions to more effectively support their open source communities.








\small
\bibliographystyle{unsrtnat}
\bibliography{orb_references}

\begin{thebibliography}{6}
\providecommand{\natexlab}[1]{#1}
\providecommand{\url}[1]{\texttt{#1}}
\expandafter\ifx\csname urlstyle\endcsname\relax
  \providecommand{\doi}[1]{doi: #1}\else
  \providecommand{\doi}{doi: \begingroup \urlstyle{rm}\Url}\fi

\bibitem[ope({\natexlab{a}})]{openfermion}
\url{https://github.com/quantumlib/OpenFermion}, {\natexlab{a}}.

\bibitem[ope({\natexlab{b}})]{openfermion-pyscf}
\url{https://github.com/quantumlib/OpenFermion-PySCF}, {\natexlab{b}}.

\bibitem[des()]{design-patterns-julia}
\url{https://github.com/PacktPublishing/Hands-on-Design-Patterns-and-Best-Practices-with-Julia}.

\bibitem[c-s()]{c-specialization}
\url{https://github.com/irenelopez42/C-specialization}.

\bibitem[vsd()]{vsdsram}
\url{https://github.com/pradeepsk13/vsdsram_sky130_1.8V}.

\bibitem[don()]{donoyo}
\url{https://github.com/donoyoyo/SMITstuff}.

\end{thebibliography}

\bibliographystyleR{unsrtnat}
\bibliographyR{repositories_minimal}

\appendix
\normalsize 

\section{Alternative Filtering Methods}
\label{app:filtering}

In addition to the approach presented in the main text, we explored several alternative filtering methods. While these methods provided useful insights, they were ultimately not used in our final analysis due to lower performance or practical limitations.

\subsubsection{\textbf{Method 1: Score Based Classifier (SBC)}}
Our baseline system is inspired by previous work on feature engineering; where a feature of the repository, such as finding a keyword associated with a university in the metadata, is assigned a score. The intuition is that whenever we detect a feature contributing to the repository’s signature of affiliation, we increase the probability that the SBC model labels it as a positive case.

In particular, to estimate the probability \( p_i \) that a repository \( r_i \) is affiliated with a university, we apply a rule-based scoring system consisting of three main steps. First, we perform a keyword search across multiple data attributes. This includes scanning the repository’s metadata component \( \text{repo}_i \) including its name, description, homepage, and README, for campus-related terms from Table~\ref{tab:uc-campuses}. We also examine the organization metadata \( \text{org}_i \). If the owner is an organization, we inspect its attributes: name, description, email, URL and company for matches. Additionally, we analyze the top two contributors \( \text{contrib}_i^{(1)} \) and \( \text{contrib}_i^{(2)} \), examining attributes such as email, name, bio, and company to determine potential university affiliation. Each attribute \( a \) has an associated probability score \( s_a \in [0,1] \) representing how likely a match in that attribute indicates university affiliation. Let:
\begin{itemize}
    \item \( m_{i,a} \in \{0, 1\} \) indicate whether repository \( r_i \) has a match in attribute \( a \),
    \item \( p_i \) be the total probability estimate for repository \( r_i \).
\end{itemize}

The total probability is computed as the sum of individual attribute probabilities, capped at 1:
\begin{equation*}
\begin{aligned}
p_i = \min \Bigg(
& \sum_{a \in \text{repo}_i} s_a m_{i,a}
+ \sum_{a \in \text{org}_i} s_a m_{i,a} \\
& + \sum_{j=1}^{2} \sum_{a \in \text{contrib}_i^{(j)}} s_a m_{i,a},
\; 1
\Bigg)
\end{aligned}
\end{equation*}

The specific scores are heuristically tuned using high-confidence attributes and are detailed in Table~\ref{tab:repo-score-table}.


\begin{table}[htbp]
\centering
\scriptsize
\renewcommand{\arraystretch}{1.2}
\begin{tabularx}{\linewidth}{|l|
  >{\hsize=1.0\hsize\raggedright\arraybackslash}X|
  >{\hsize=1.0\hsize\raggedright\arraybackslash}X|
  >{\hsize=1.0\hsize\raggedright\arraybackslash}X|}
\hline
\textbf{Metadata} & \textbf{Attribute Checked (\( a \))} & \textbf{Matched Criteria} & \textbf{Probability Score (\( s_a \))} \\
\hline
\multirow{2}{*}{\( \text{repo}_i \)} 
& \texttt{homepage} & University domain & 1.0 \\
\cline{2-4}
& \texttt{readme}, \texttt{description}, \texttt{name} & University name, acronym, or alt. queries & 0.20 per match per attribute \\
\hline
\multirow{3}{*}{\( \text{org}_i \)} 
& \texttt{url} & University domain & 1.0 \\
\cline{2-4}
& \texttt{email} & University domain & 1.0 \\
\cline{2-4}
& \texttt{name}, \texttt{description}, \texttt{company} & University name, acronym, or alt. queries & 0.30 per match per attribute \\
\hline
\multirow{2}{*}{\( \text{contrib}_i^{(j)} \)} 
& \texttt{email}, \texttt{name}, \texttt{bio}, \texttt{company} & University domain & 0.50 per match per attribute \\
\cline{2-4}
& \texttt{email}, \texttt{name}, \texttt{bio}, \texttt{company} & University name, acronym, or alt. queries & 0.20 per match per attribute \\
\hline
\end{tabularx}
\caption{Scoring system for estimating university affiliation.}
\label{tab:repo-score-table}
\end{table}

\subsubsection{\textbf{Method 2: Text Embeddings \& Machine Learning}}


Instead of relying only on keyword matches, our next approach creates a feature vector by using an embedding with the repository's signature, and then uses that feature vector as an input to a machine-learning classifier. Notice that the embedding captures the broader context in which university-related terms appear. For example, the phrase “developed by UC Santa Cruz” carries much stronger affiliation than an incidental mention of the campus name.

\noindent \textbf{Embedding:} We aggregate the repository attributes into a single text block. For each repository \( r_i \), we define: \\
$
\mathbf{m}_i = \left( \text{repo}_i, \text{org}_i, \text{contrib}_i^{(1)}, \text{contrib}_i^{(2)} \right),
$
where each component is a set of textual fields. We concatenate these fields into a single raw text string \( t_i \in \mathbb{T} \), such that $t_i = \text{concat}(\mathbf{m}_i).$ We finally transform each \( t_i \) into a vector representation \( \mathbf{v}_i \in \mathbb{R}^d \) using OpenAI's text embedding model \cite{openai2024embeddings} \( \mathcal{E} : \mathbb{T} \rightarrow \mathbb{R}^d \).  Given the input size constraints of the small embedding model, we truncate each text field to a maximum of 20,000 characters. Since the README is typically the longest field, this truncation primarily affects its content.

\noindent \textbf{Training:} We randomly sample 200 repositories and manually assign a binary label to each one, \( y_i \in \{0, 1\} \), where \( y_i = 1 \) indicates university affiliation. We intentionally avoid balancing the dataset to preserve the natural class distribution.

We tested several machine learning models, including neural networks, random forests, and Support Vector Machines (SVM). The classification results were very similar among all models, but SVMs gave slightly better results, so we showcase them for the rest of this paper. 
After training, for each repository, the model outputs a probability per class:
$
p_i = f_{\text{prob}}(\mathbf{v}_i),
$
where \( p_i \) is the predicted probability that \( r_i \) is affiliated with the target university.

This approach requires a manually labeled training dataset, which introduces significant overhead in terms of human effort and time.

\section{Cost and Runtime Comparison}
\label{app:cost}
Given the complexity of detecting affiliated repositories, it is also important to consider the practical trade offs in cost and runtime across different modeling approaches. Table \label{tab:model-costs-times-236k} shows a cost comparison for different methods of processing GitHub repository data. For the SVM approach the cost reflects transforming the text into vector embeddings. The total time is computed by adding the embeddings time to the model training and prediction time, which in most cases is around 0.70 seconds. For the other methods, costs are based on using the OpenAI API with prompts whose length includes both dynamic repository data and a static prompt component. We compute average input sizes by concatenating all relevant data fields, resulting in approximately 2,900 characters of dynamic text plus about 1,600 characters of static prompt text for the API models. The cost estimates are scaled to 236K repositories, which corresponds to the total number of repositories found for the UC System. The time is computed for \texttt{gpt-4o} and \texttt{gpt-3.5} For \texttt{gpt-5-mini}, which is much slower, we report the single-threaded runtime estimated from multithreaded benchmarks.

While \texttt{gpt-5-mini} achieves competitive classification accuracy and is relatively inexpensive compared to \texttt{gpt-4o}, its runtime presents a significant practical limitation for large-scale datasets. As shown in Table~\ref{tab:model-costs-times-236k}, processing 236,000 repositories with \texttt{gpt-5-mini} costs just over \$100, making it a cost-effective option. However, the single-threaded runtime is estimated to exceed 21 days, reflecting its much slower processing speed compared to \texttt{gpt-4o} or \texttt{gpt-3.5}. These results illustrate a key tradeoff between cost, accuracy, and efficiency, emphasizing that model selection should consider not only predictive performance but also practical constraints such as dataset size and processing time.

\makeatletter
\@dblfloatplacement
\makeatother
\begin{table*}[!t]
\centering
\scriptsize
\begin{tabular}{cccccccc
}
\toprule
\makecell{\textbf{Model}} & 
\makecell{\textbf{Avg.} \\ \textbf{Chars}} & 
\makecell{\textbf{Input} \\ \textbf{Tokens}} & 
\makecell{\textbf{Output} \\ \textbf{Tokens}} & 
\makecell{\textbf{Price per 1M} \\ \textbf{Tokens (I/O)}} & 
\makecell{\textbf{Cost per} \\ \textbf{Repo}} & 
\makecell{\textbf{Cost ×} \\ \textbf{236K}} & 
\makecell{\textbf{Time} \\ \textbf{(min)}} \\
\midrule
SBC        & —     & —     & —    & \$0 / \$0             & \$0         & \$0        & 18       \\
SVM        & 2,900 & 725   & —    & \$0.00002 / —         & \$0.0000145 & \$3.42     & 1022     \\
gpt-3.5    & 4,500 & 1,125 & 100  & \$0.0005 / \$0.0015   & \$0.00056 / \$0.00015 & \$167.6    & 3708     \\
gpt-4o     & 4,500 & 1,125 & 100  & \$2.50 / \$10.00      & \$0.00381   & \$899.75   & 4330     \\
gpt-5-mini & 4,500 & 1,125 & 100  & \$0.25 / \$2.00       & \$0.00048   & \$113.46   & 30342    \\
\bottomrule
\end{tabular}
\caption{Cost and average time comparison for 236,000 inputs using OpenAI models, based on average input size and token pricing. Single-threaded time is used for gpt-5-mini.}
\label{tab:model-costs-times-236k}
\end{table*}

\end{document}